\documentclass[preprints,article,accept,moreauthors,pdftex]{mdpi} 

\firstpage{1}
\pubvolume{xx}
\issuenum{1}
\articlenumber{5}
\pubyear{2020}
\copyrightyear{2020}
\history{Received: date; Accepted: date; Published: date}
\usepackage[utf8]{inputenc}
\usepackage{amsmath}
\usepackage{amsthm}
\usepackage{amsfonts}
\usepackage{amssymb}
\usepackage{mathtools}
\usepackage{caption}
\usepackage{afterpage}
\DeclareGraphicsExtensions{.pdf,.png,.jpg,.mps,.eps,.ps}

\usepackage{subcaption}
\usepackage{rotating}

\usepackage{alphalph}

\usepackage{forloop}
\newcommand{\defvec}[1]{\expandafter\newcommand\csname v#1\endcsname{{\mathbf{#1}}}}
\newcounter{ct}
\forLoop{1}{26}{ct}{
	\edef\letter{\alph{ct}}
	\expandafter\defvec\letter
}

\forLoop{1}{26}{ct}{
	\edef\letter{\Alph{ct}}
	\expandafter\defvec\letter
}

\Title{Birhythmic Analog Circuit Maze:\\
A Nonlinear Neurostimulation Testbed}
\Author{Ian D. Jordan $^{1,2}$ and Il Memming Park $^{1,2,3}$\orcidB{}}

\AuthorNames{Ian D. Jordan and Il Memming Park}

\address{%
$^{1}$ \quad Department of Applied Mathematics and Statistics, Stony Brook University, Stony Brook, NY, USA\\ %; ian.jordan@stonybrook.edu\\
$^{2}$ \quad Institute for Advanced Computing Science,
 Stony Brook University, Stony Brook, NY, USA\\
$^{3}$ \quad Department of Neurobiology and Behavior, Stony Brook University, Stony Brook, NY, USA}

\corres{Correspondence: memming.park@stonybrook.edu}

\abstract{
Brain dynamics can exhibit narrow-band nonlinear oscillations and multistability.
For a subset of disorders of consciousness and motor control, we hypothesize that some symptoms originate from the inability to spontaneously transition from one attractor to another.
Using external perturbations, such as electrical pulses delivered by deep brain stimulation devices, it may be possible to induce such transition out of the pathological attractors.
However, the induction of transition may be non-trivial, rendering the current open-loop stimulation strategies insufficient.
In order to develop next-generation neural stimulators that can intelligently learn to induce attractor transitions, we require a platform to test the efficacy of such systems.
To this end, we designed an analog circuit as a model for the multistable brain dynamics.
The circuit spontaneously oscillates stably on two periods as an instantiation of a 3-dimensional continuous-time gated recurrent neural network.
To discourage simple perturbation strategies such as constant or random stimulation patterns from easily inducing transition between the stable limit cycles, we designed a state-dependent nonlinear circuit interface for external perturbation.
We demonstrate the existence of nontrivial solutions to the transition problem in our circuit implementation.
}


\begin{document}
%%%%%%%%%%%%%%%%%%%%%%%%%%%%%%%%%%%%%%%%%%

\section{Introduction}

Multistability is a widespread phenomenon in the field of dynamical systems where a system exhibits multiple stable states or more generally attractors~\cite{pisarchik_control_2014}.
Appearing in nearly all disciplines of natural science and engineering, including 
biology \cite{abou-jaoude_mechanisms_2011,decroly_birhythmicity_1982}, 
chemistry \cite{hudson_chaos_1981,ngonghala_extreme_2011},  
electronics \cite{biswas_control_2016,in_experimental_2003}, 
fluid mechanics \cite{ravelet_multistability_2004,shiau_multistability_1999},
genetics \cite{koseska_inherent_2007,labavic_networks_2014}, and
physics \cite{ding_interaction_2004,kiss_chaotic_2003}, 
researchers have shown substantial interest in such behavior \cite{pisarchik_control_2014}.
In neuroscience, computations are thought to be implemented by multistable dynamical systems, and recent experimental and methodological advances have generated renewed interests~\cite{little1974,breakspear_dynamic_2017,fairhall_editorial_2017,sussillo_neural_2014,La_Camera2019,Zhao2016d}.
The multiple attractors within these dynamical systems seem to underlie a wide array of functions, including sensory perception~\cite{li_attention_2017}, motor function~\cite{churchland_neural_2012}, and cognition~\cite{mante_context-dependent_2013,jazayeri_navigating_2017}, as well as dysfunctions such as movement disorders~\cite{Wagenaar1996}, epilepsy~\cite{Iasemidis2003}, and disorders of consciousness~\cite{Lewis2018,Schiff2014}.
We hypothesize that multistability underlie some dynamical neurological diseases such that manifested symptoms are fundamentally due to the inability to naturally transition from one basin of attraction to another.
Under this hypothesis, neurostimulation techniques provide a means to perturb neural systems to assist transition between attractors as a treatment option.

Open-loop electrical stimulation therapies have shown remarkable successes, most notably with Parkinson's disease~\cite{little_focusing_2014}.
However, open-loop strategies are likely to be insufficient for the general induction of attractor transitions that manifest complex nonlinear dynamics and non-trivial stimulus induced perturbations.
For example, high-amplitude low-frequency signals, such as those dominant in disorders of consciousness~\cite{Schiff2014}, suggest the existence of strong attractor dynamics which may require a sophisticated feedback control system to transition out~\cite{hocker_myopic_2019}.
This sets the stage for the next-generation closed-loop neural stimulators that can intelligently learn to induce attractor transitions.
The added complexity of the closed-loop stimulation systems calls for a platform to develop and test their efficacy.

In this paper, we aim to develop a hardware platform by which these intelligent stimulation algorithms can be tested and validated on. As is a common method for realizing a dynamical system physically, an analog electronic circuit exhibiting the desired dynamics is constructed~\cite{bao_numerical_2017,rahman_aminur_qualitative_2018,zhao_streaming_2019}. Due to the nature of widely used electrical stimulation for neurological implants, this medium will serve well as a testbed.
To reduce unnecessary complexity, we construct our system to demonstrate the simplest form of oscillatory multistability, birhythmicy. More specifically, the system will simultaneously exhibit two self-exciting limit cycles of notably different frequencies. Given that the state of the system is sufficiently close to one of the two attractors, an intelligent stimulation algorithm can be tested by trying to perturb the system into the other basin of attraction.

In the following section, we derive the system from the general continuous-time dynamical system underlying the gated recurrent unit (GRU), a commonly used recurrent neural network architecture~\cite{cho_learning_2014,jordan_gated_2019}.
In section~\ref{section:circuit}, we discuss the details of the circuit design and present the results of the physical realization.
In section~\ref{section:stimulator}, we discuss the addition and design of a nonlinear circuit, state dependent on the system described in sections \ref{section:dynsys} and \ref{section:circuit}, by which external stimulation is interfaced.
The addition of this nonlinear \textit{stimulator circuit} will ensure random or periodic stimulation patterns will be ineffective in inducing transitions between the two attractor states.

\section{Birhythmic Dynamics in 3-dimensions}\label{section:dynsys}
Our goal is to find a simple bistable dynamical system where each attractor corresponds to a periodic orbit.
We draw from the recurrent neural network literature on simple forms of stable limit cycles.
Specifically, we utilize the autonomous continuous-time gated recurrent unit (ct-GRU)~\cite{jordan_gated_2019,cho_learning_2014} formulation, which can be represented as follows:
\begin{align} \label{eq:GRU1}
    \dot{\vh} &= (1 - \vs(t)) \odot (\vT(t) - \vh(t))
    \qquad & \text{(hidden state)} &
\\ \label{eq:GRU2}
    \vs(t) &= \sigma(\vU_s \vh(t) + \vb_s)
    \qquad & \text{(update gate)} &
\\	\label{eq:GRU3}
    \vr(t) &= \sigma(\vU_r \vh(t) + \vb_r)
    \qquad & \text{(reset gate)} &
\\  
    \vT(t) &= \tanh(\vU_h(\vr(t) \odot \vh(t)) + \vb_h)
	   & &
\label{eq:GRU}
\end{align}
where $\vh(t) \in \mathbb{R}^d$ is the state of the system, $\vU_s, \vU_r, \vU_h \in \mathbb{R}^{d \times d}$ are the parameter matrices, $\vb_s, \vb_r, \vb_h \in \mathbb{R}^{d}$ are the bias vectors, $\odot$ represents the Hadamard product, and $\sigma(\vz) = {1}/{(1 + e^{-\vz})}$ is the element-wise logistic sigmoid function.
For a given set of parameters, fixed points of the system exist where $\dot{\vh} = 0$.
Since $1 - \vs(t) > 0, \, \forall\vs$, this term does not influence the roots of the right-hand side of \eqref{eq:GRU1}. 
Therefore, $\vs(t)$ can only affect the speed of the flow, and in turn can be neglected when choosing a set of parameters for the system to enact a desired structure of attractors \cite{jordan_gated_2019}.
Note that if the parameters of $\vr(t)$ have been set to zero, the ct-GRU architecture simplifies to the classic ct-tanh-RNN if the parameters of $\vs(t)$ are also set to zero.

\begin{figure}[t!h!b]
	\centering
	\includegraphics[width=0.75\textwidth]{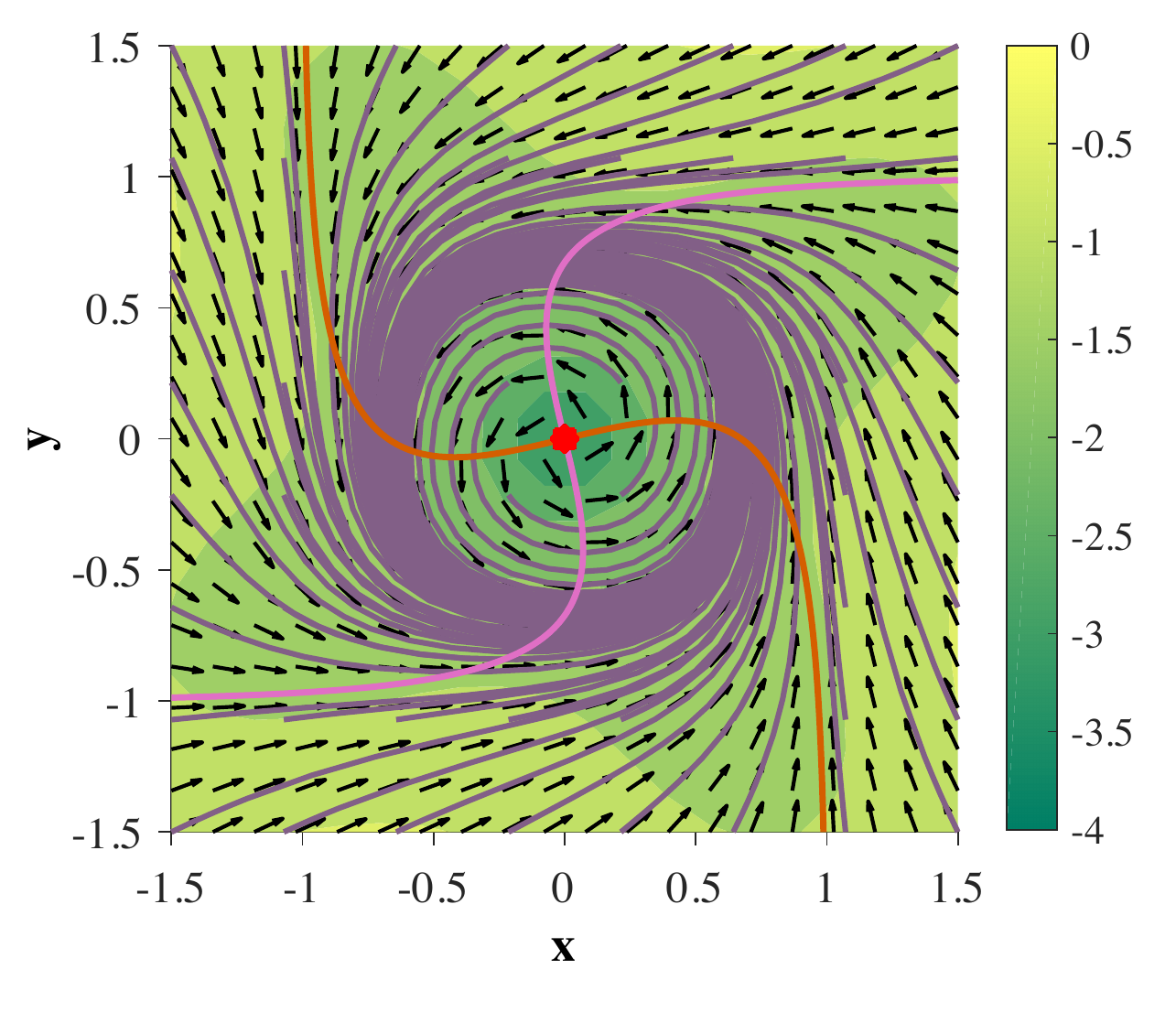}
	\caption{
		\textbf{Planar Limit Cycle with 2D ct-GRU depicted in phase space:}
		the red dot indicates an unstable fixed point at the origin unstable, orange and pink lines represent the x and y nullclines respectively. Purple lines indicate various trajectories of the hidden state. Direction of the flow is determined by the black arrows, where the colormap underlying the figure depicts the magnitude of the velocity of the flow in log scale. 
	}
	\label{fig:limitCyclePlanar}
\end{figure}

In previous work~\cite{jordan_gated_2019}, we have shown that for $d = 2$ the ct-GRU is capable of expressing a single limit cycle (attracting closed orbit) in phase space under the following set of parameters:
\begin{equation} \label{eq:2Dparam}
\vU_r,\vb_r,\vb_h = 0,\; 
\vU_h = 3\begin{bmatrix}
\cos{\alpha} & -\sin{\alpha} \\
\sin{\alpha} & \cos{\alpha}
\end{bmatrix}
\end{equation}
where $\alpha \in S_0$ and $S_0 \supseteq (\frac{\pi}{21},\frac{\pi}{3.8})$.
The phase portrait depicting this behavior for $\alpha = \frac{\pi}{5}$ can be seen in Fig.~\ref{fig:limitCyclePlanar}, where $\vh \equiv \begin{bmatrix}
x & y
\end{bmatrix}^T$.

\begin{figure}[t!h!b]
	\centering
	\includegraphics[width=0.6\textwidth]{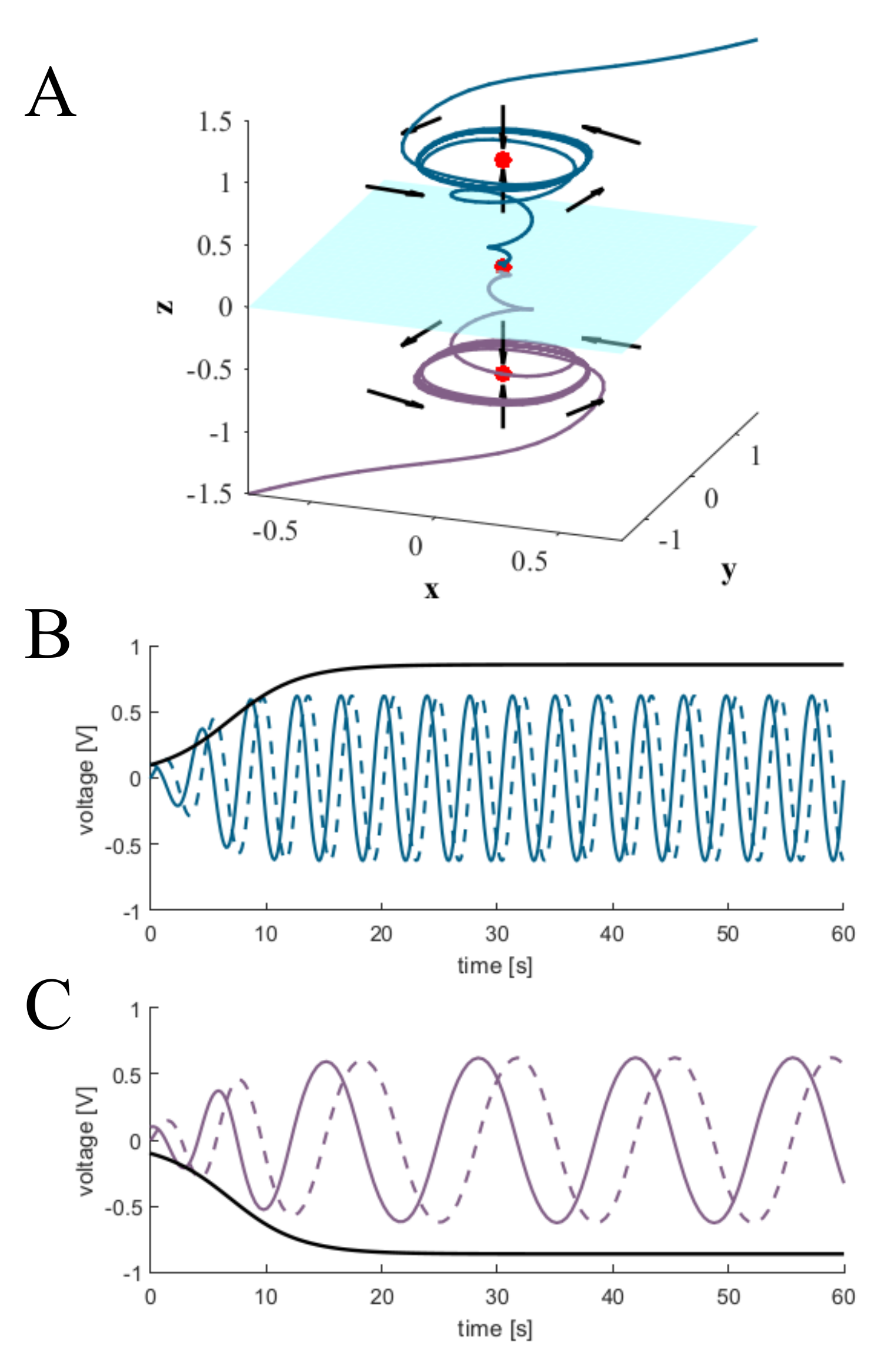}
	\caption{
		\textbf{Birhythmicy in 3-dimensions:}
		(A): light blue manifold on the $x-y$ plane separates the basins of attraction of the upper and lower limit cycles. Trajectories are colored either dark blue or purple depending on which basin of attraction they are initialized in. Red dots indicate fixed points, and black arrows depict the direction of flow. (B),(C): $x$, $y$, and $z$ components of trajectories initialized in the basins of attractions for the top and bottom limit cycles respectively. Solid colored lines indicate $x(t)$, dashed lines indicate $y(t)$, and black lines indicate $z(t)$.
	}
	\label{fig:limitCycle3D}
\end{figure}

Extending the system to 3-dimensions allows for the simultaneous existence of two limit cycles in phase space under a single set of parameters. More specifically, the addition of a third dimension enables us to mirror any attractor structure representable for $d = 2$ across an unstable manifold on the plane defined by the original two dimensions in $\mathbb{R}^3$.
This behavior is depicted in Fig.~\ref{fig:limitCycle3D}(A), where now $\vh \equiv \begin{bmatrix}
x & y & z
\end{bmatrix}^T$, and the parameters are set as follows:
\begin{equation} \label{eq:3Dparam}
\vU_r, \vb_r, \vb_h = 0,\; 
\vU_h = 3\begin{bmatrix}
\cos{\frac{\pi}{5}} & -\sin{\frac{\pi}{5}} & 0 \\
\sin{\frac{\pi}{5}} &  \cos{\frac{\pi}{5}} & 0 \\
0 &             0 & 1 
\end{bmatrix}
\end{equation}

As stated before, $\vs(t)$ only acts to adjust the speed of phase flow. If $\vU_s, \vb_s = 0$ the periods of both limit cycles are equal. As a means to easily decouple the two frequencies of oscillation, the velocity of flow may be made dependent on its vertical position with respect to the $z$-axis.
While the range of the logistic sigmoid function has the benefit of always being defined on $(0,1)$, it may produce inaccurate results when physically realized along its tails.
Furthermore, any function that is strictly positive and sufficiently well-behaved on the region of phase space we're interested in will work under this context.

For simplicity, we redefine $\vs(t)$ linearly as $\vs(t) = \vU_s \vh(t) + \vb_s$.
We then note that~\eqref{eq:GRU} is asymptotically bound to $[-1,1]^d$. To account for potential error in the electronic realization, we set $\vs(t)$ such that its output remains strictly positive on $(-1.5,1.5)\times(-1.5,1.5)\times\mathbb{R}$. The results of this linear update-gate are demonstrated in Fig.~\ref{fig:limitCycle3D}(B) and \ref{fig:limitCycle3D}(C), and achieved under the following set of parameters in conjunction with those of \eqref{eq:3Dparam}:
\begin{equation} \label{eq:zparam}
\vb_s = \begin{bmatrix}
-0.5\\
-0.5\\
0.5
\end{bmatrix},
\vU_s = \begin{bmatrix}
0 & 0 & 1 \\
0 & 0 & 1 \\
0 & 0 & 0 
\end{bmatrix}
\end{equation}
To better grasp the dynamical system depicted in Fig.~\ref{fig:limitCycle3D} to be later realized, we can rewrite \eqref{eq:GRU1}--\eqref{eq:GRU} explicitly in terms of $x,y$ and $z$ with our chosen parameters from \eqref{eq:3Dparam} and \eqref{eq:zparam} as follows:
\begin{align}
\dot{x} &=
    \left(z - \frac{3}{2}\right)
    \left[
	x - \tanh\left(
		x \cdot \frac{3}{2}\cos\frac{\pi}{5} -
		y \cdot\frac{3}{2}\sin\frac{\pi}{5}
    \right)\right]
\label{eq:system1}
\\
\dot{y} &= 
    \left(z - \frac{3}{2}\right)
    \left[
	y - \tanh\left(x \cdot \frac{3}{2}\sin\frac{\pi}{5}
	+ y \cdot \frac{3}{2}\cos\frac{\pi}{5}
    \right)\right]
\label{eq:system2}
\\
\gamma\dot{z} &= 
    -\frac{1}{2}
    \left[z - \tanh\left(\frac{3}{2}z\right)\right]
\label{eq:system3}
\end{align}
where $\gamma \in \mathbb{R}$ is an added time constant that will be implemented in the circuit realization to adjust the difficulty of transitioning between attracting states. For our implementation of \eqref{eq:system1}--\eqref{eq:system3} we let $\gamma = 10^6$.

\section{Electronic Physical Realization} \label{section:circuit}
Within most applications, smooth continuous-time systems can be realized as electronic circuits comprised of inexpensive components and integrated circuits \cite{bao_numerical_2017}. In this section we introduce a comprehensive circuit design to realize \eqref{eq:system1}--\eqref{eq:system3} and construct the system on a breadboard. Experimental recordings of trajectories of interest are then compared with the theoretical system derived in section \ref{section:dynsys} as a means to validate the realization. All basic operational amplifiers used are TL082CP and all individual transistors are MPS2222. In addition, two analog multiplier chips are used, which are the standard AD633 four quadrant multipliers. Note that all schematics shown assume unity gain associated with each multiplier. A complete list of all component values in the following schematics can be found in appendix \ref{appendixA}. 

\subsection{Nonlinear activation function circuit}
\begin{figure}[t!h!b]
	\centering
	\includegraphics[width=0.7\textwidth]{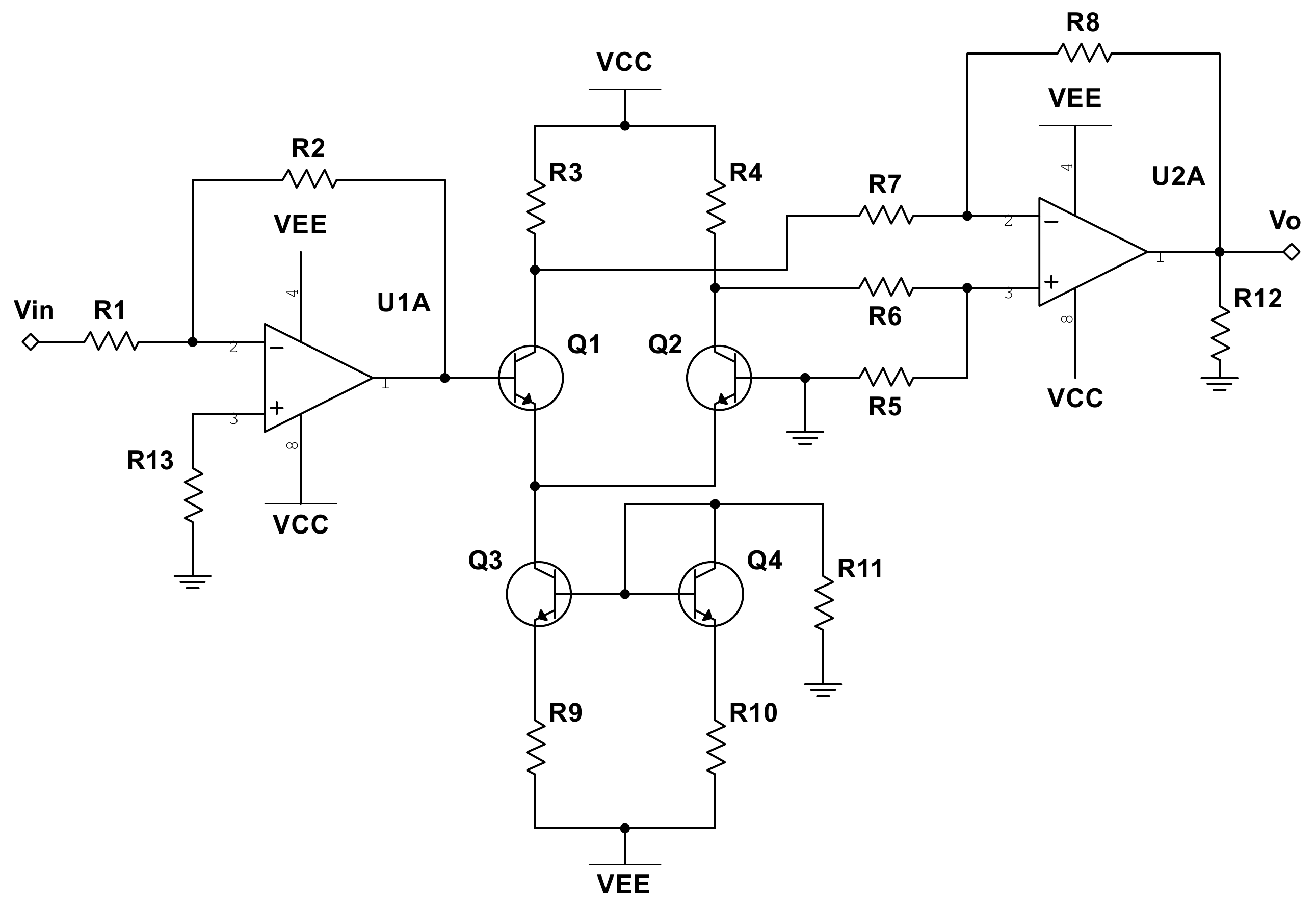}
	\caption{Electronic circuit realization of the hyperbolic tangent function, as implemented in \cite{bao_numerical_2017}. $V_{in}$ and $V_o$ represent the input and output signals respectively.
	}
	\label{fig:tanhCircuit}
\end{figure}

To properly realize \eqref{eq:system1}--\eqref{eq:system3} as an analog circuit, we first must account for the nonlinearity in the system; the hyperbolic tangent function. Previous work has allowed us to easily realize this nonlinearity by means of a simple op-amp and transistor circuit as depicted in Fig.~\ref{fig:tanhCircuit}~\cite{bao_numerical_2017}.
Allowing $V_{in}$ and $V_{o}$ to represent the input and output voltages of the circuit respectively, \citet{duan_electronic_2007} showed that the input-output relation takes the following form:
\begin{equation} \label{eq:inoutrelation}
V_0 = -\tanh
    \left(
	\frac{R_2}{2R V_T}V_{in}
    \right)
\end{equation}
where $V_T \approx 26$ mV is the thermal voltage of the transistors at room temperature. Allowing  $R_2 = 520\Omega$, $R_3 = R_4 = 1k\Omega$, $R_{11} = 11k\Omega$, all other resistors set to $R=10k\Omega$, $V_{CC} = 15$V and $V_{EE} = -15$V, reduces the coefficient in front of $V_{in}$ in \eqref{eq:inoutrelation} to 1, thereby successfully implementing the hyperbolic tangent function as an analog circuit. Further information regarding the error associated with our constructed hyperbolic tangent units can be found in appendix \ref{appendixB}.
\afterpage{\clearpage}

\subsection{Schematics of electronic birhythmic RNN}
Analog circuits can successfully perform addition and subtraction with just operational amplifiers and appropriately connected resistors. The additional use of capacitors allows for an analog implementation of integration \cite{wilmshurst_analog_2001}. In regards to our realization of \eqref{eq:system1}--\eqref{eq:system3}, the analog implementation of the hyperbolic tangent function can be achieved with the schematic demonstrated in Fig.~\ref{fig:tanhCircuit}.
In the following schematics such nonlinear operations are represented by boxes labeled ``\texttt{-tanh}'', where terminals IOP1 and IOP2 are the input and output voltage to each hyperbolic tangent unit respectively. Furthermore, two analog multiplication chips can be used to introduce the phase flow speed dependence on $z(t)$ in \eqref{eq:system1} and \eqref{eq:system2}. Thus the entire system can be constructed entirely from simple analog components.

\begin{figure}[t!h!b]
	\centering
	\includegraphics[width=1\textwidth]{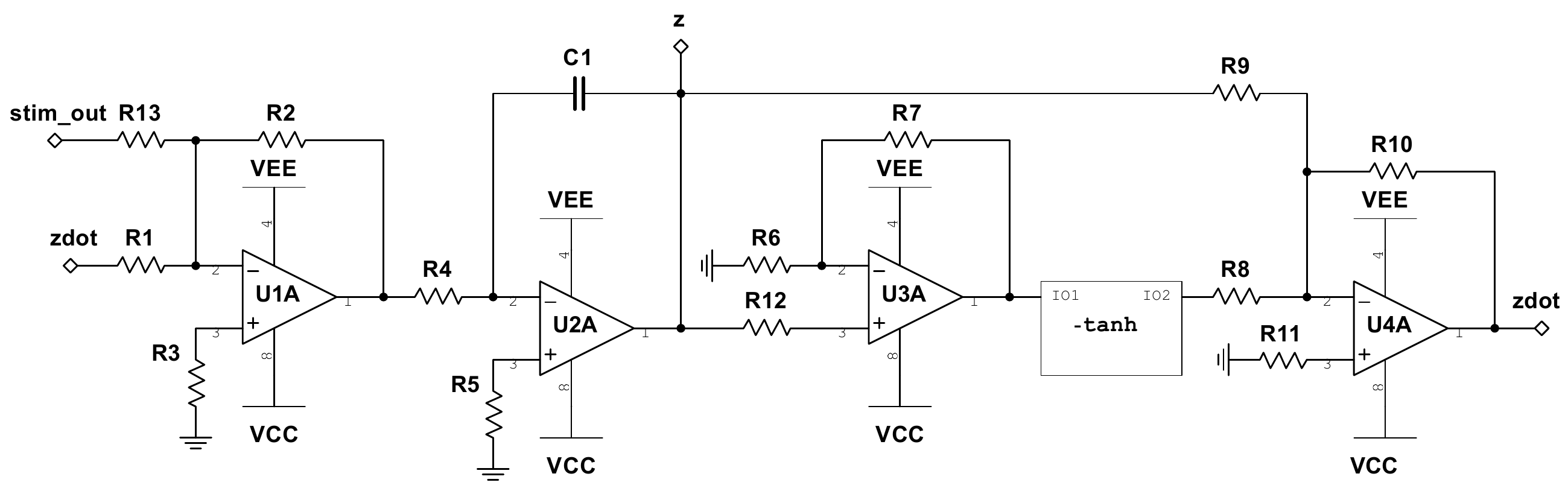}
	\caption{Circuit schematic of $\dot{z}$ for the birhythmic system. The system block labeled \texttt{-tanh} represents the circuit depicted in Fig. \ref{fig:tanhCircuit}, where \texttt{I01} and \texttt{I02} correspond to $V_{in}$ and $V_o$ respectively. The terminal labeled \texttt{stim\_out} represents the output to the stimulator circuit, as discussed in section \ref{section:stimulator}.
	}
	\label{fig:z}
\end{figure}

\begin{figure*}[t!h!b]
	\centering
	\includegraphics[width=1\textwidth]{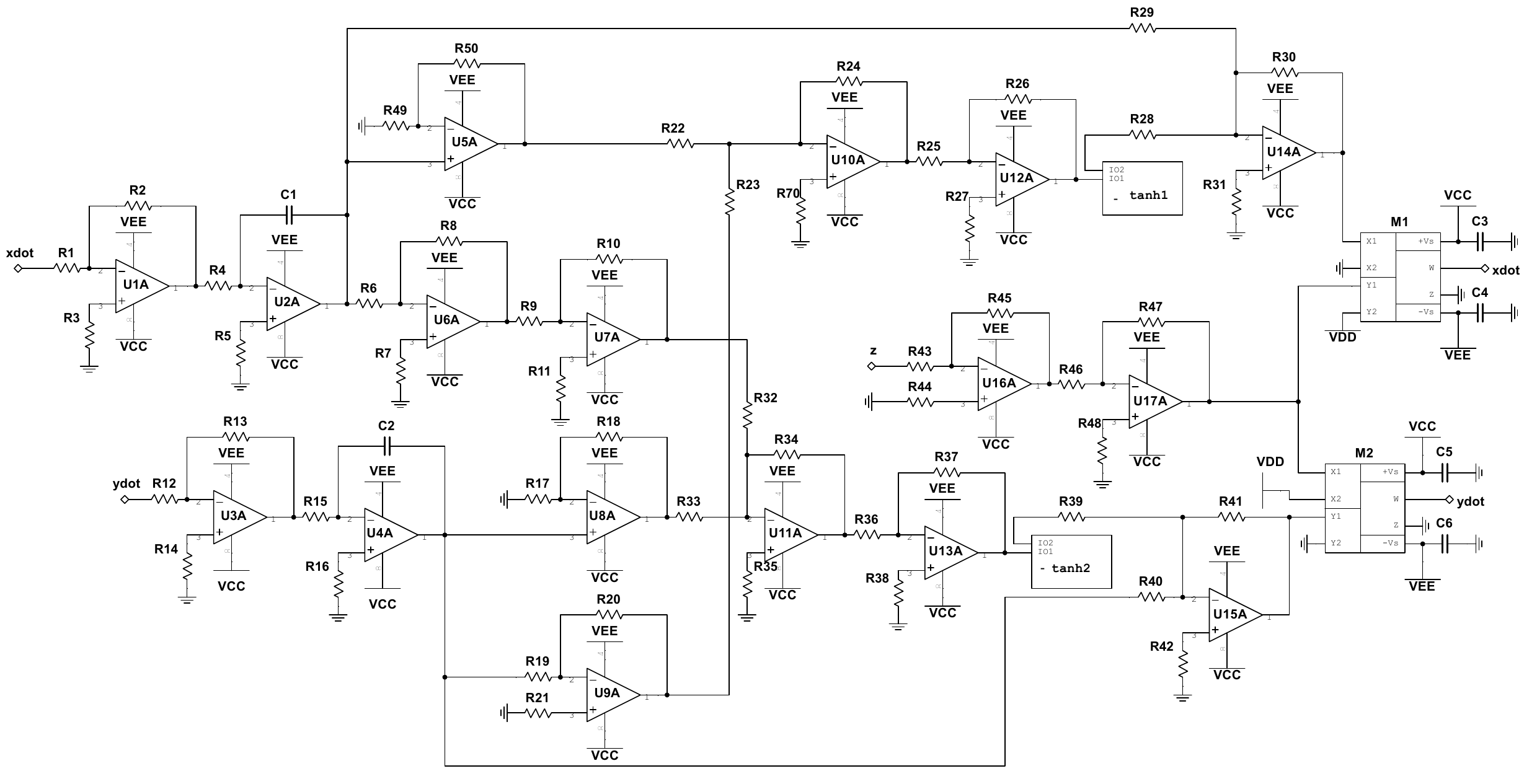}
	\caption{Circuit schematic of $\dot{x}$ and $\dot{y}$ for the birhythmic system. The system blocks labeled \texttt{-tanh1} and \texttt{-tanh2} represent the circuit depicted in Fig.~\ref{fig:tanhCircuit}, where \texttt{I01} and \texttt{I02} correspond to $V_{in}$ and $V_o$ respectively for both blocks.
	The two multiplier chips M1 and M2 are assumed to operate with unity gain.
	}
	\label{fig:xy}
\end{figure*}

We begin with the circuit realization of \eqref{eq:system3}, where the schematic is shown in Fig. \ref{fig:z}. Note that \eqref{eq:system3} only comprises of the state variable $z(t)$, and can therefore be build independently of \eqref{eq:system1} and \eqref{eq:system2}. The variable $z(t)$ is represented as the voltage across capacitor C1 in the provided circuit diagram. The input reading \texttt{stim\_out} represents the output of the nonlinear stimulator circuit, and will be discussed in section \ref{section:stimulator}.
The schematic of the electronic realization of \eqref{eq:system1} and \eqref{eq:system2} is shown in Fig. \ref{fig:xy}, where the state variables, $x(t)$ and $y(t)$, are represented by the voltages across capacitors C1 and C2 respectively. The ``z'' input to analog multipliers, M1 and M2, is taken from the integrator output labeled ``z'' in Fig. \ref{fig:z}.

\subsection{Circuit construction and experimental results}

\begin{figure}[t!h!b]
	\centering
	\includegraphics[width=0.9\textwidth]{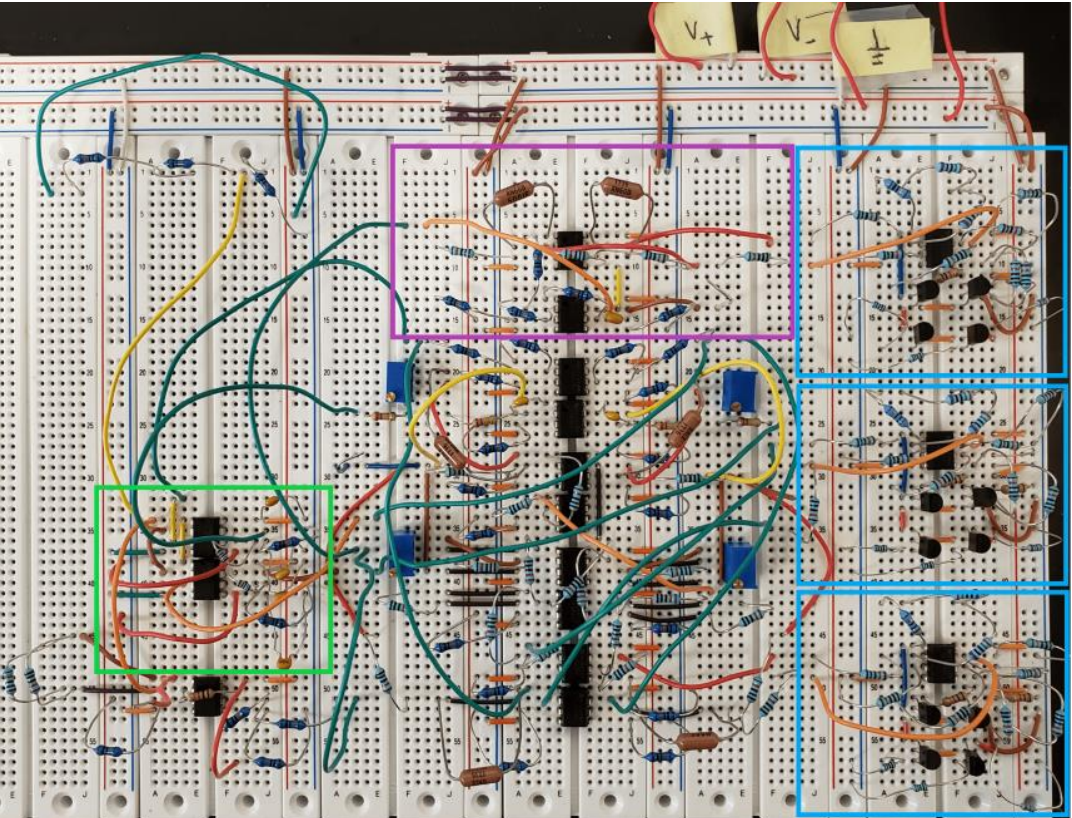}
	\caption{Physical birhythmic circuit constructed on a breadboard. Blue Boxes represent hyperbolic tangent units. The magenta box indicates the subsection of the circuit generating $\dot{z}$, and the green box indicates the analog multipliers.
	}
	\label{fig:physicalCircuit}
\end{figure}

A circuit following the schematics depicted in Fig.~\ref{fig:z} and Fig.~\ref{fig:xy} was constructed on a breadboard. This system was photographed and is depicted in Fig.~\ref{fig:physicalCircuit}. The blue boxes indicate the three separate hyperbolic tangent units. The magenta box indicates the realization of \eqref{eq:system3}, as described by Fig.~\ref{fig:z}, and the green box highlights the two analog multiplier chips and their configuration.
\afterpage{\clearpage}

\begin{figure}[t!h!b]
	\centering
	\includegraphics[width=0.9\textwidth]{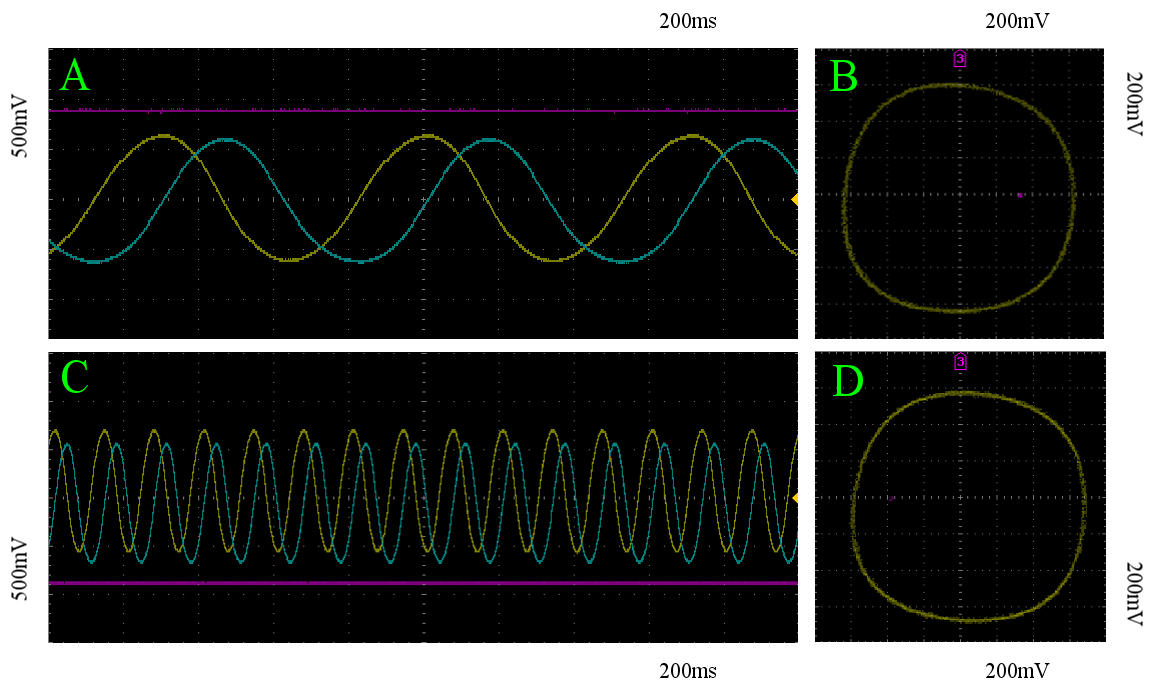}
	\caption{\textbf{Experimental recordings of the birhythmic circuit:} (A),(C): $x$ (yellow), $y$ (blue), and $z$ (pink) with respect to time of trajectories within the basin of attraction of the fast and slow limit cycles respectively. (B),(D): Projection of the corresponding trajectories in (A) and (C) onto the x-y plane respectively.
	}
	\label{fig:experimentalPlots}
\end{figure}

Using a four channel oscilloscope, we observed the behavior of the system for different initial voltages across the capacitors. These recordings are depicted in Fig.~\ref{fig:experimentalPlots} for two different initializations; one within the basin of attraction for each of the two stable limit cycles. Fig.~\ref{fig:experimentalPlots}(A) and \ref{fig:experimentalPlots}(C) show the asymptotic behavior of the three state variable voltages in time $[x(t), y(t), z(t)]$.
As intended, the period of the each limit cycles are visibly different, and qualitatively match the asymptotic behavior demonstrated in Fig.~\ref{fig:limitCycle3D}(B) and \ref{fig:limitCycle3D}(C). Furthermore, Fig. \ref{fig:experimentalPlots}(B) and \ref{fig:experimentalPlots}(D) depict the trajectories projected onto the x-y plane, shown in Fig. \ref{fig:experimentalPlots}(A) and \ref{fig:experimentalPlots}(C) respectively. These recordings indicate the same asymptotic behavior as demonstrated in Fig. \ref{fig:limitCyclePlanar}. As such, we can conclude that the analog implementation of \eqref{eq:system1}--\eqref{eq:system3} was successfully realized from the model derived in section \ref{section:dynsys}.
\afterpage{\clearpage}

\section{Nonlinear Stimulator Circuit Design} \label{section:stimulator}
When looking at the system derived in section \ref{section:dynsys}, as expressed by \eqref{eq:system1}--\eqref{eq:system3}, we notice the geometric simplicity of the global dynamics.
As shown in Fig.~\ref{fig:limitCycle3D}(A), the two periodic attractors are mirror images of one another across a planar unstable manifold on the x-y plane. This symmetry enabled us to easily decouple the frequencies of each limit cycle by introducing a strictly positive variable time constant dependent only on our $z$-coordinate. However, this introduces a clear problem from the point of view of developing a neurostimulation testbed. In order to transition between basins of attraction, one need only worry about the component $z(t)$.
In other words, transition between attractors can be achieved with constant stimulation on $z$.
This solution to attractor transition is trivial, and does not require the use of an intelligent algorithm to solve, rendering it inadequate as a testbed.
In order to negate this issue, we develop a state-dependent nonlinear stimulator interface circuit by which all stimulus must pass through.
In addition, we allow only one location of stimulation within the circuit previously designed, as marked by the node labelled \texttt{stim\_out} in Fig.~\ref{fig:z}, where the output to the stimulator circuit will be fed in and summed with the current value of $\dot{z}$ in the system.
By extending the system properly in this way, we can prevent straightforward stimulation patterns (i.e constant, random, periodic, etc.) from inducing attractor transitions.

\begin{figure}[t!h!b]
	\centering
	\includegraphics[width=0.5\textwidth]{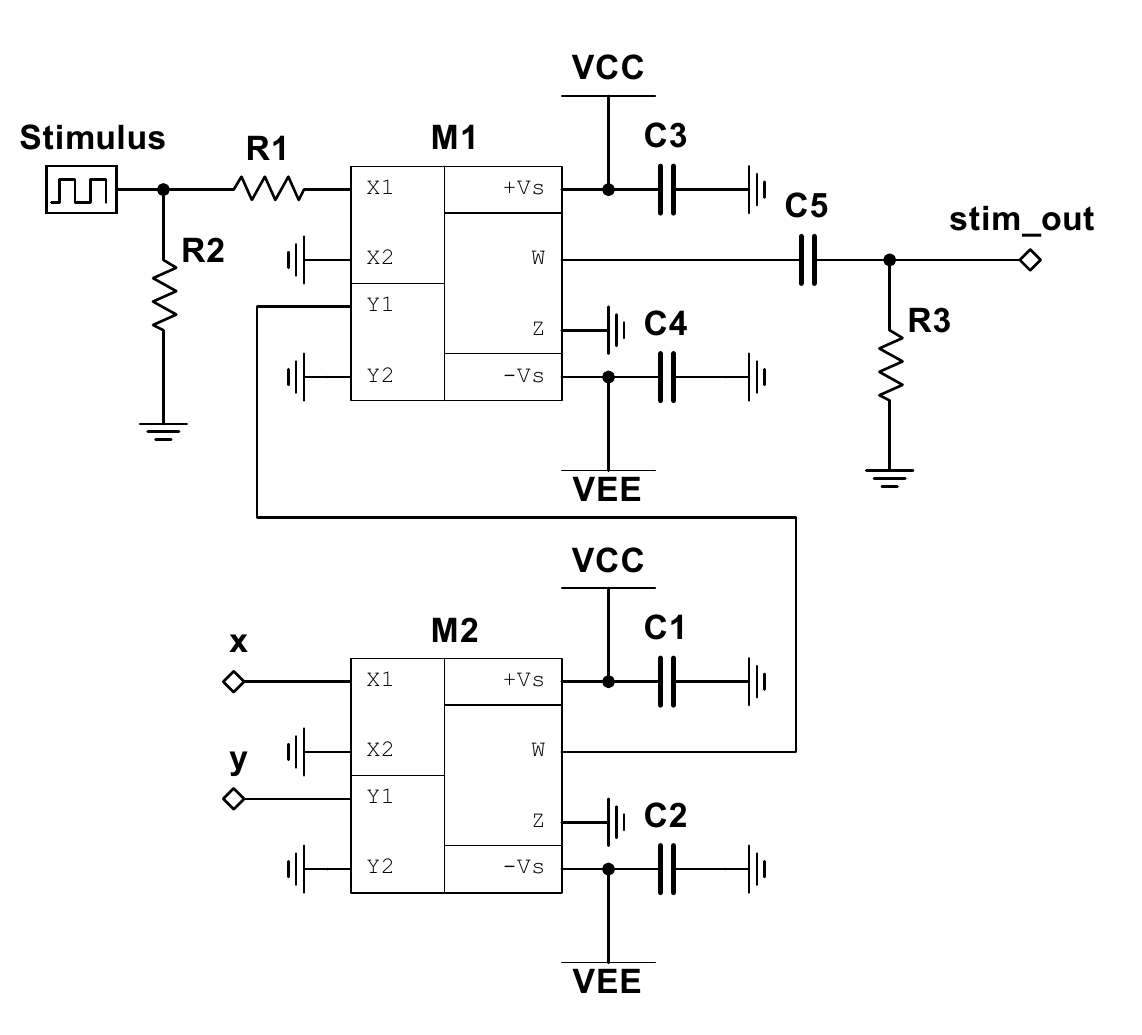}
	\caption{Schematic for nonlinear \textit{stimulator circuit}, with input labeled as \textit{Stimulus}. The output, labeled \texttt{stim\_out}, is fed into the terminal with the same name presented in the circuit diagram shown in Fig. \ref{fig:z}. The two multiplier chips M1 and M2 are assumed to operate with unity gain, and the $x$ and $y$ terminals are fed into the equivalently named terminals depicted in Fig. \ref{fig:xy}.
	}
	\label{fig:stimulator}
\end{figure}

Ideally, we want to develop a stimulation circuit such that a pulse stimulated at a random time my cause the output, \texttt{stim\_out}, to either increase or decrease. Furthermore, the amount by which the signal can change should exist on a continuum, rather than outputting a voltage from a finite set of values. The final requirement we will enforce for such a circuit will be that a stimulation pulse delivered at a random time should have equal probability of increasing or decreasing the output signal. This last requirement ensures that if one stimulates randomly or continuously, the expected value of \texttt{stim\_out} averaged across all time will be zero as time approaches infinity.

Due to the sinusoidal nature of the $x$ and $y$ components in~\eqref{eq:system1} and \eqref{eq:system2}, \eqref{eq:stim} will serve as the input/output relation of the stimulator circuit at a given time $t$.
\begin{equation} \label{eq:stim}
    S_{out}(t) = S_{in}(t) x(t) y(t)
\end{equation}
where $S_{out}$ is the output voltage of the stimulator circuit, $S_{in}$ is the input stimulus (Note that $S_{in} = 0$ when no stimulus is applied). We note that this system satisfies all of our requirements, as $z$ is independent of $x$ and $y$. To transition from the slow limit cycle to the fast limit cycle, stimulation must be applied primarily when $x(t)$ and $y(t)$ are of opposite signs. Similarly, to transition from the fast limit cycle to the slow limit cycle, stimulation must be applied primarily when $x(t)$ and $y(t)$ are of the same sign. However, since the speed of oscillation is dependent on $z(t)$, any added stimulation will change the time window by which proper stimulation should be applied, ensuring that a periodic stimulation regime will fail to transition between the attractors.
Fig.~\ref{fig:stimulator} depicts the schematic for the described stimulator circuit, where $S_{in}$ is labeled as \texttt{Stimulus}.

\begin{figure}[t!h!b]
	\centering
	\includegraphics[width=0.8\textwidth]{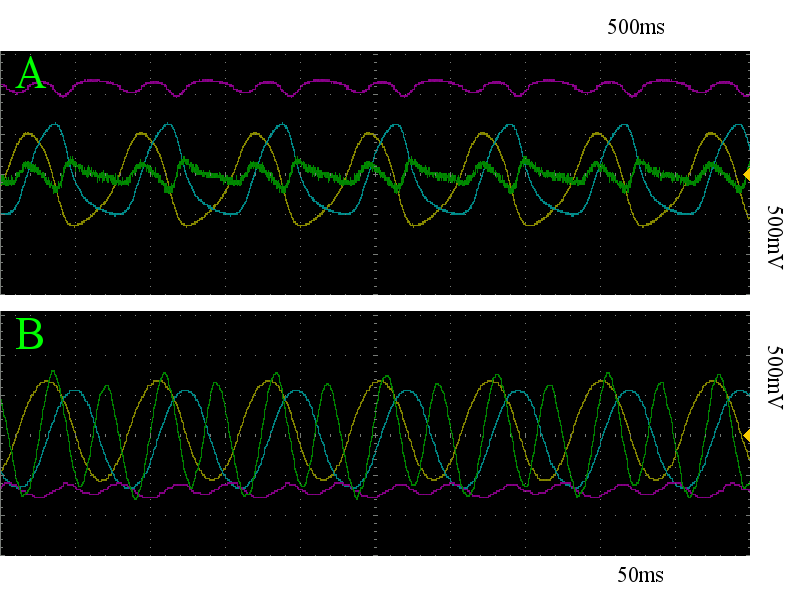}
	\caption{x (yellow), y (blue), z (pink), and \textit{stim\_out} (green) with respect to time of trajectories within the basin of attraction of the slow (A) and fast (B) limit cycles, under constant stimulation. Note that this stimulation regime cannot successfully transition between the two attracting states in either direction.}
	\label{fig:constantStim}
\end{figure}

\begin{figure}[t!h!b]
	\centering
	\includegraphics[width=0.8\textwidth]{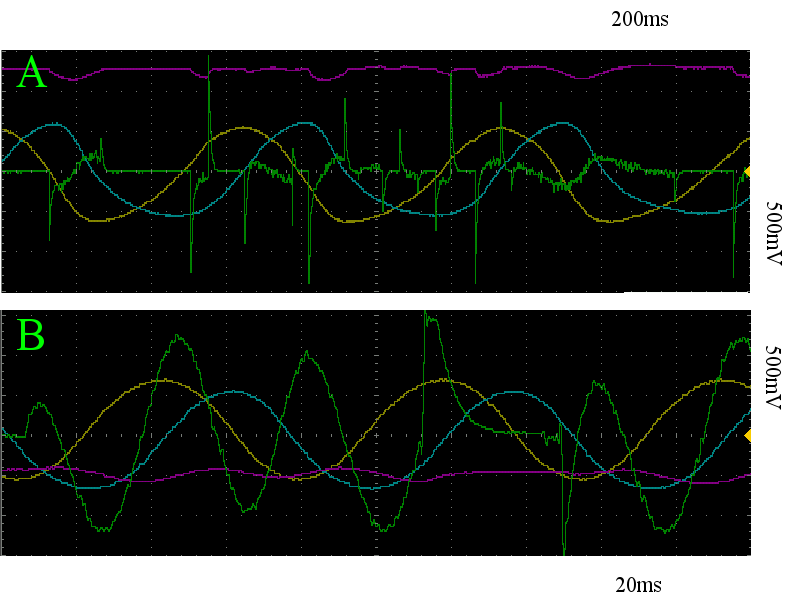}
	\caption{x (yellow), y (blue), z (pink), and \textit{stim\_out} (green) with respect to time of trajectories within the basin of attraction of the slow (A) and fast (B) limit cycles, with random stimulation. Note that this stimulation regime does not successfully transition between the two attracting states in either direction.}
	\label{fig:randomStim}
\end{figure}

A physical realization of the stimulator circuit was created in conjunction with the birhythmic system and tested with a stimulus that could be turned on (4V) or off (0V). Figures \ref{fig:constantStim}(A) and \ref{fig:constantStim}(B) depicts the resultant behavior of the system under constant 4V stimulation, initialized on the slow and fast limit cycles respectively. Again, the yellow, blue, and pink curves represent the $x$, $y$, and $z$ components of the system, and the green curve depicts the voltage over time of \texttt{stim\_out}.
Similarly, figures \ref{fig:randomStim}(A) and \ref{fig:randomStim}(B) depict the same system under randomly timed manual stimulation.
As expected, both of these stimulation regimes do not escape the basin of attraction by which the system is initialized in.
This is because a transition between basins of attraction would require the coincidental stimulation when the system state is in the proper two quadrants of the x-y plane to overtake the stimulation in the other two for a prolonged period of time, which is unlikely.

\begin{figure}[t!h!b]
	\centering
	\includegraphics[width=0.8\textwidth]{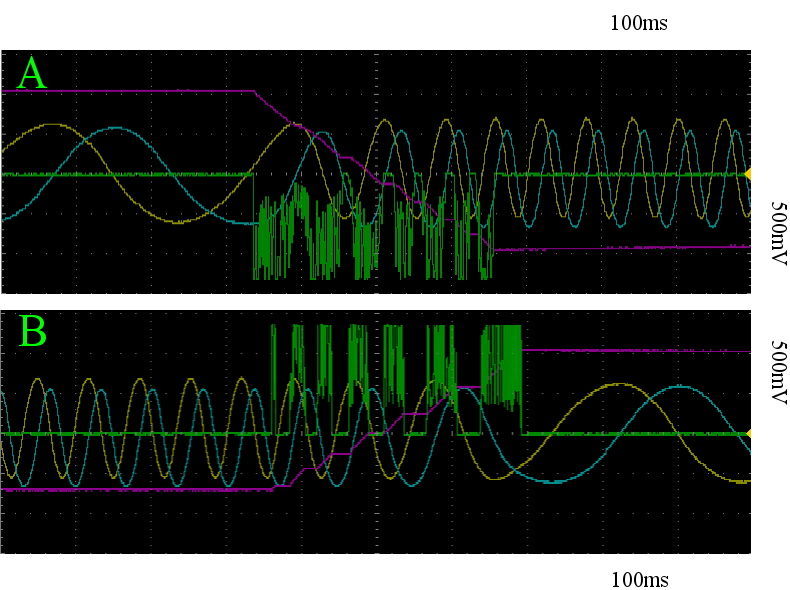}
	\caption{Solution stimulation pattern to transition between states. x (yellow), y (blue), z (pink), and \textit{stim\_out} (green) with respect to time of trajectories initialized within the basin of attraction of the slow (A) and fast (B) limit cycles. As $z(t)$ changes with stimulation so does the frequency of oscillation. As such, the time window for successful stimulation changes with each pulse.}
	\label{fig:solutionStim}
\end{figure}

Finally, we have implemented a solution pattern that can successfully transition between the stable limit cycle attractors.
We have simply placed a rectifier diode in the appropriate orientation just prior to the \texttt{stim\_out} node in Fig.~\ref{fig:stimulator}, and applied constant stimulation.
The results are shown in Fig.~\ref{fig:solutionStim}.
Such a demonstration indicates the existence of a stimulation pattern capable of transitioning between basins of attraction in the system that could be mimicked without the added rectifier with the properly learned input stimulus.

\afterpage{\clearpage}

Note that a neurostimulation system would not have direct access to the state variables, the dynamical system, nor how the stimulus modulates the states.
Any such algorithm will need to discover the latent dynamics and learn to control the states from observations~\cite{zhao_streaming_2019} at the same time it learns to transition out of the current basin of attraction.

\section{Conclusion}
In this paper, an electronic testbed for intellegent neurostimulation methods is developed from a physical realization of the dynamical system underlying the architecture of an artificial gated recurrent neural network. Using simple analog components, the system is fabricated such that it exhibits birhythmic behavior, but the stimulation pattern required to transition between attracting states is made nontrivial. As such, standard open-loop stimulation regimes will be unable to induce attractor transitions, thereby enabling the system to validate the efficacy of next-generation neurostimulation algorithms upon sucessfully jumping between basins of attraction.  

%%%%%%%%%%%%%%%%%%%%%%%%%%%%%%%%%%%%%%%%%%
%\setcounter{section}{-1} %% Remove this when starting to work on the template.

%%%%%%%%%%%%%%%%%%%%%%%%%%%%%%%%%%%%%%%%%%
\authorcontributions{I.J and I.M.P conceptualized the project. I.J. conducted experiments. I.J and I.M.P wrote the manuscript. All authors have read and agreed to the published version of the manuscript.}

%%%%%%%%%%%%%%%%%%%%%%%%%%%%%%%%%%%%%%%%%%
\funding{
This work is supported by Stony Brook University's Discovery Award, NIH EB-026946, and NSF IIS-1845836. I.J was supported partially by the Institute for Advanced Compuational Science Jr. Researcher Fellowship, at Stony Brook University.
}

%%%%%%%%%%%%%%%%%%%%%%%%%%%%%%%%%%%%%%%%%%
\conflictsofinterest{The authors declare no conflict of interest.} 

%%%%%%%%%%%%%%%%%%%%%%%%%%%%%%%%%%%%%%%%%%
%% optional
\appendixtitles{yes} %Leave argument "no" if all appendix headings stay EMPTY (then no dot is printed after "Appendix A"). If the appendix sections contain a heading then change the argument to "yes".
\appendix
\section{Circuit component values}\label{appendixA}
\unskip
\subsection{Component values for Fig. \ref{fig:z}}
\noindent$VCC = 15V$, $VEE = -15V$
\\
$R1 = R3 = R4 = R5 = R6 = R12 = R13 = 100k\Omega$, $R2 = 1M\Omega$, $R7 = 50k\Omega$, $R8 = R9 = R11 = 10k\Omega$, $R10 = 5k\Omega$
\\
$C1 = 1\mu F$

\subsection{Component values for Fig. \ref{fig:xy}}

\noindent$VCC = 15V$, $VEE = -15V$, $VDD = 1V$
\\
$R1 = R3 = R5 = R6 = R7 = R9 = R10 = R11 = R12 = R14 = R16 = R17 = R19 = R21 = R22 = R23 = R24 = R25 = R26 = R27 = R28 = R29 = R31 = R32 = R33 = R34 = R35 = R36 = R37 = R38 = R39 = R40 = R42 = R43 = R44 = R45 = R46 = R47 = R48 = R49 = 10k\Omega$, $R2 = R4 = R13 = R15 = 100k\Omega$, $R8 = R20 = 8820\Omega$, $R18 = R50 = 2315\Omega$, $R30 = R41 = 5k\Omega$.
\\
$C1 = C2 = 10 \mu F$, $C3 = C4 = C5 = C6 = 1\mu F$

\subsection{Component values for Fig. \ref{fig:stimulator}}

\noindent$VCC = 15V$, $VEE = -15V$
\\
$R1= R2 = R3 = 100k\Omega$
\\
$C1 = C2 = C3 = C4 = C5 = 1\mu F$

\section{Hyperbolic tangent implementation}\label{appendixB}

\begin{figure}[thb]
	\centering
	\includegraphics[width=0.45\textwidth]{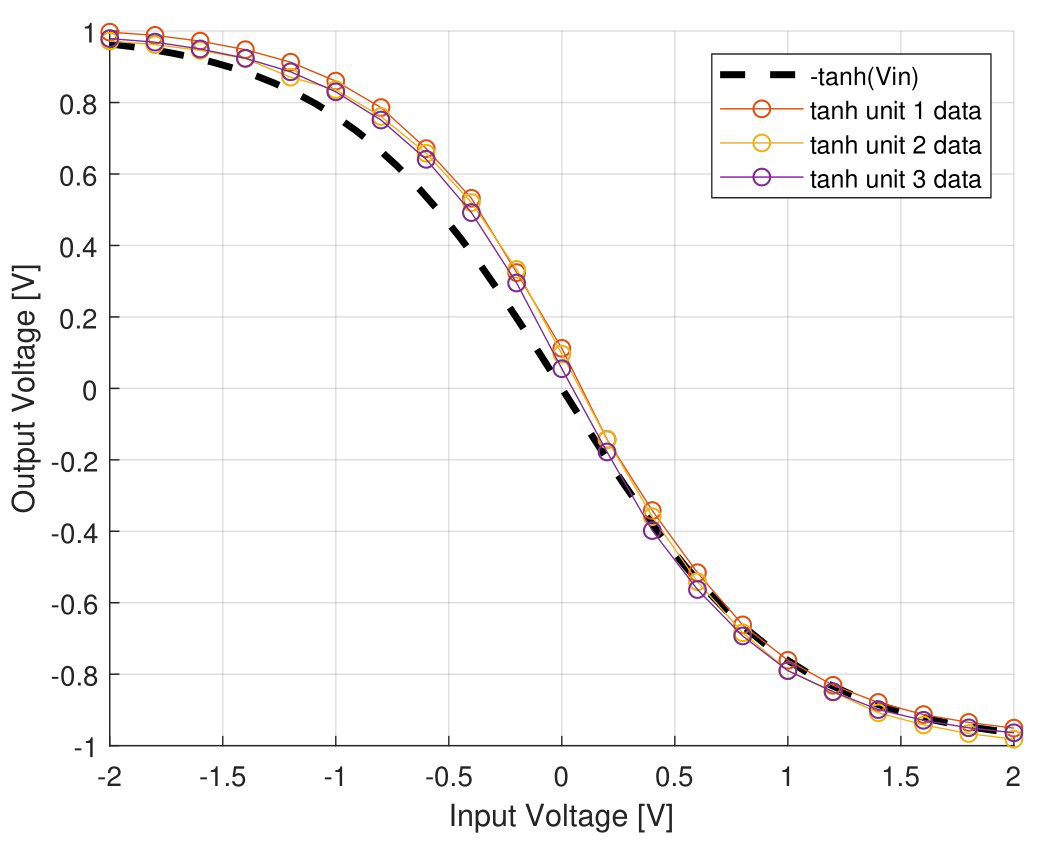}
	\caption{Input/output relation of three physically realized hyperbolic tangent circuits, interpolated through 21 points, compared with the analytic hyperbolic tangent function.
	}
	\label{fig:tanherror}
\end{figure}

Our system depicted in \eqref{eq:system1}--\eqref{eq:system3} requires the construction of three separate hyperbolic tangent units.
Following construction, 21 input voltages, equally spaced on [-2,2]V, were applied to each hyperbolic tangent unit as input, and each respective output voltage was recorded and compared with the hyperbolic tangent function for that given input. These recordings and comparisons of accuracy of the three hyperbolic tangent units are displayed in Fig. \ref{fig:tanherror}. The average mean squared error of the built hyperbolic tangent units with respect to the hyperbolic tangent function was approximately 86.5 mV.

%%%%%%%%%%%%%%%%%%%%%%%%%%%%%%%%%%%%%%%%%%
\reftitle{References}

\externalbibliography{yes}
\bibliography{bistableBIB}

%%%%%%%%%%%%%%%%%%%%%%%%%%%%%%%%%%%%%%%%%%
\end{document}